\documentclass[twocolumn]{aastex63}

\usepackage{newtxtext,newtxmath}
\usepackage[T1]{fontenc}
\usepackage{ae,aecompl}

\usepackage{graphicx}	% Including figure files
\usepackage{amsmath}	% Advanced maths commands
\usepackage{amssymb}	% Extra maths symbols
\usepackage{nicefrac}
\usepackage{subfig}

\received{XXX}
\revised{YYY}
\accepted{ZZZ}
\submitjournal{ApJ}

\shorttitle{AIC and ejected mass}
\shortauthors{Sharon \& Kushnir}

\begin{document}
%%%%%%%%%%%%%%%%%%% TITLE PAGE %%%%%%%%%%%%%%%%%%%

% Title of the paper, and the short title which is used in the headers.
% Keep the title short and informative.
\title{Towards an accurate description of an accretion-induced collapse and the associated ejected mass}

% The list of authors, and the short list which is used in the headers.
% If you need two or more lines of authors, add an extra line using \newauthor

\correspondingauthor{Amir Sharon}
\email{amir.sharon@weizmann.ac.il}

\author{Amir Sharon}
\affiliation{Department of Particle Physics \& Astrophysics, Weizmann Institute of Science, Rehovot 76100, Israel}

\author{Doron Kushnir}
\affiliation{Department of Particle Physics \& Astrophysics, Weizmann Institute of Science, Rehovot 76100, Israel}

% Enter the current year, for the copyright statements etc.
%\pubyear{2019}

% Don't change these lines

% Abstract of the paper
\begin{abstract}
We revisit the accretion-induced collapse (AIC) process, in which a white dwarf collapses into a neutron star. We are motivated by the persistent radio source associated with the fast radio burst FRB 121102, which was explained by Waxman as a weak stellar explosion with a small ($\sim 10^{-5}M_{\odot}$) mildly relativistic mass ejection that may be consistent with AIC. Additionally, the interaction of the relatively low ejected mass with a pre-collapse wind might be related to fast optical transients. The AIC is simulated with a one-dimensional, Lagrangian, Newtonian hydrodynamic code. We put an emphasis on accurately treating the equation of state and the nuclear burning, which is required for any study that attempts to accurately simulate AIC. We leave subjects such as neutrino physics and general relativity corrections for future work. Using an existing initial profile and our own initial profiles, we find that the ejected mass is $\sim 10^{-2}$ to $10^{-1}M_{\odot}$ over a wide range of parameters, and we construct a simple model to explain our results.
\end{abstract}

% Select between one and six entries from the list of approved keywords.
% Don't make up new ones.

.

%%%%%%%%%%%%%%%%%%%%%%%%%%%%%%%%%%%%%%%%%%%%%%%%%%

%%%%%%%%%%%%%%%%% BODY OF PAPER %%%%%%%%%%%%%%%%%%

\section{Introduction}

accretion-induced collapse (AIC) is a theorized process in which a white dwarf (WD) collapses into a neutron star (NS), followed by an explosion that ejects a fraction of the star's mass at mildly relativistic velocities. AIC has been proposed as the outcome of an accreting WD and as a possible NS formation channel \citep{canal1980r,nomoto1986fate}. Existence of young pulsars in globular clusters \citep{lyne1996psr} suggests that some fraction of NSs indeed were created in this way. A renewed interest in this process has arisen recently, following new astronomical discoveries that may be related \citep{Waxman2017,lyutikov2018fbots,moriya2019vtc}.

One discovery that may be related to AIC is the persistent radio source associated with FRB 121102 \citep{scholz2016repeating}. The fast radio burst (FRB) source resides in a dwarf galaxy at a distance of $\sim970\,\text{Mpc}$, with a persistent radio source located in that direction \citep{chatterjee2017direct}. \citet{Waxman2017} suggested that the persistent radio source was created by a weak stellar explosion, with a small ($\sim 10^{-5}M_{\odot}$) mildly relativistic mass ejection, which may be consistent with AIC, while the resulting NS acted as the source of the FRBs \citep[see also][]{margalit2019fast,kashiyama2017testing}.

Another window for observing AIC might be fast-rising blue optical transients \citep{drout2014rapidly}. The interaction of the ejected mass with a pre-collapse wind could be related to these events \citep{lyutikov2018fbots}, which are characterized by a short optical rise time, $<10\,\text{days}$, and a peak luminosity comparable to supernovae. For analysis of the expected light curve of AIC, see \cite{Metzger2009} and \cite{darbha2010nickel}.

The AIC process is the consequence of a WD, which is being held mostly by electron degeneracy pressure, that accumulates enough mass to surpass the Chandrasekhar mass limit, $M_\text{Ch}$, at which point it can no longer support its own mass and collapses. The collapse should happen without a prior thermonuclear runaway that will explode the star before it is able to collapse. The conditions that determine the fate of the WD were studied, for example, by \citet{nomoto1986fate} and \citet{Tauris2013}. In this work we assume that a nuclear runaway did not occur and that the WD was able to collapse. In this case, the energies gained by nuclear reactions ($\sim\text{MeV/baryon}$) are not sufficient for a nuclear runaway, since the gravitational binding energy is much deeper, as shown below.
The WD collapses over its own freefall time, $t_{ff}\sim0.1\,\text{s}$, until the core reaches nuclear densities ($\rho_{nuc}\sim10^{14}\,\text{g}\:\text{cm}^{-3}$) and repelling strong interactions halt the collapse. The radius of the core at bounce is
\begin{equation}
    R\approx\left(\frac{3}{4\pi}\frac{M_\text{Ch}}{\rho_{nuc}}\right)^{1/3}\cong 20\,\text{km},
\end{equation}
roughly a factor of $5$ larger than the Schwarzschild radius, $r_s=2GM_\text{Ch}/c^2\cong 4\,\text{km}$. The characteristic velocity of the ejected material is comparable to the escape velocity, $v_{esc}=\sqrt{r_s/R}c\cong 0.5c$, and is mildly relativistic.
% The escape velocity from the core is therefore $v_{esc}=\sqrt{r_s/R}c\sim0.5c$, which will be the characteristic velocities of the ejected material, and are mildly relativistic.
The gravitational energy per baryon around the core is
 \begin{equation}
    \epsilon\sim\frac{3GM_\text{Ch}}{5R}\sim50\,\text{MeV/baryon},
\end{equation}
which is converted to internal and kinetic energy, corresponding to a temperature of $T\lesssim 5\times10^{11}\,\text{K}$. During the collapse, the outer layers bounce off the dense core, creating a shock wave that propagates outward, ejecting some mass in the process. Since most of the gravitational energy is converted to internal energy, only some fraction of the initial mass will be ejected, while the remaining core becomes an NS. 

%\textbf{In order for AIC to successfully occur, the WD should approach the Chandrasekhar mass and collapse without a thermonuclear runaway explosion. The conditions that will determine the fate of the WD were studied, for example, by \citet{nomoto1986fate} and \citet{Tauris2013}. In this work we assume that a nuclear runaway did not occur and that the WD is able to collapse.}

Simulating AIC raises some challenges that should be carefully addressed. The range of densities and temperatures throughout the process varies greatly, from nuclear densities ($\rho_{nuc}\sim10^{14}\,\text{g}\:\text{cm}^{-3}$) at core bounce to very low densities ($<10^5\,\text{g}\:\text{cm}^{-3}$) at the edge of the star. The entire range should be described accurately and smoothly by the equation of state (EOS). The high temperatures and densities cause nuclear reactions to be very rapid, and these reactions should be treated in a self-consistent way. Neutrinos play an important part in the process \citep{woosley1992collapse}, 
%. Matter is being deleptonized, creating and eventually emitting neutrinos in the process. Also, neutrino anti-neutrino pairs are thermally created in these regimes of densities and temperatures. 
and general relativity (GR) corrections should also be accounted for, as the radius of the core after bounce is comparable to the Schwarzschild radius, leading to alterations of the proto-neutron star \citep[PNS][]{liebendorfer2001}. 

Several previous studies have estimated the amount of ejected mass from an AIC. \citet{woosley1992collapse} estimated an ejected mass of $M_\text{ej}\sim0.01M_{\odot}$, using a progenitor made by \citet{nomoto1986fate}, and one-dimensional (1D) hydrodynamic simulations that include neutrino transport. They used the EOS that was developed by Baron, Cooperstein, and Kahana \citep[BCK;][]{baron1985type}, and no GR corrections were included. Matter was flash burned to nuclear statistical equilibrium (NSE) when the temperature exceeded $5\times10^9\,\text{K}$. In their simulations, the prompt bounce shock failed to explode the star, and the mass loss occurred through neutrino-driven wind. \citet{Fryer1999} used the same initial profile and the EOS of \cite{herant1994}, which couples the Lattimer-Swesty (LS) nuclear EOS \citep{lattimer1991generalized} with a low-density EOS, and found that the prompt bounce shock ejects most of the mass. They obtained ejecta mass of $M_\text{ej}\sim0.1-0.3\,M_{\odot}$ for different variants of the LS EOS, nuclear burning assumptions, neutrino physics assumptions, and GR effects. They showed that the main difference with the work of \citet{woosley1992collapse} was the different EOS, where the results from the BCK EOS never agree with the results from the LS EOS, despite a wide range of attempted physical parameters. This was also the case in \cite{Swesty1994ApJ}, who compared the two EOSs and argued that the BCK EOS is less accurate. \citet{Fryer1999} further compared the obtained low-$Y_e$ ejected mass with the abundance of heavy elements in the Galaxy to derive an upper limit on the AIC rate in the Galaxy of $\sim10^{-5}\,\text{yr}^{-1}$. \citet{Dessart2006} performed multidimensional simulations and found an ejected mass of $M_\text{ej}\sim10^{-3}M_{\odot}$. They used the HShen EOS \citep{Shen1998}, extended to low densities and temperatures, and included neutrino emission and transport, but they did not include nuclear burning or GR corrections. The ejection was through a successful shock followed by a neutrino-driven wind. \cite{Dessart2006} suggested that the discrepancy with the previous works is because of the multidimensional treatment and the high-rotating progenitor. %However, the different hydrodynamic schemes and EOS might be a significant source of uncertainty in these studies and one of the reasons for the different results. 

\subsection{Objectives}

A careful treatment of the EOS is an essential part in core-collapse supernova simulations, since the dynamics is determined by the behavior of matter at nuclear densities \citep{bethe1990supernova,da2017new}. The same holds also for the AIC process, although not studied in a comparable detail to the core-collapse supernova problem. A significant complication for a calculation of the AIC problem is that the edge of the star must be included in the simulation, in which the density drops significantly, and thus an accurate description of both the high- and low-density regimes, as well as the intermediate regime that connect them, is required. For this purpose, the preparation of a relevant EOS for several regimes (of density, temperature, composition) is necessary, and the different regimes should be smoothly connected. Since only a small fraction of the star is being ejected, the intermediate regimes, in which matter can change its density and temperature by several orders of magnitudes during the collapse, play an important role in determining the amount of ejected mass. In this work we introduced and applied a scheme that smoothly and accurately connects the different regions, in a manner that enables every part of the star to be described by the relevant EOS at any time. There are two types of transitions that should be taken into account. The first occurs when the matter starts from having some composition of isotopes, and as it heats up during the collapse and undergoes nuclear reactions, the composition can reach NSE. The second is the transition from ideal gas to nuclear density EOS, occurring when the density rises above $ \sim10^{11}\,\text{g}\:\text{cm}^{-3}$.
Previous works did not treat self-consistently both transitions. It was either that nuclear reactions were neglected (or matter instantly converted to NSE state when the temperature crossed some threshold) or that the transition to nuclear density EOS was not smooth \citep{herant1994}.

Nuclear burning is also important in order to have a star with appropriate initial conditions that will successfully collapse in a feasible manner. Having the entire star at NSE prior to collapse is unrealistic, since the density and temperature near the edge are far below NSE conditions. Nuclear burning is required for matter originally not in NSE to have a smooth transition to NSE when the temperature is high enough. If nuclear burning is not enabled and the composition for some part of the star is fixed, the EOS will eventually fail to describe the matter when its density becomes high enough for nuclear interactions to be important ($\rho\gtrsim10^{11}\,\text{g}\:\text{cm}^{-3}$). Additionally, electron degeneracy causes the temperature to be very sensitive to the internal energy, resulting in large fluctuations during the collapse. This issue was also handled in our work (see Appendix \ref{app:entropy}), in a manner that has not been done in previous AIC studies.

We aim in this paper to provide a numerical calculation of an AIC (for given initial conditions and input physics) for the spherical case, where neutrinos and GR corrections are neglected. We demonstrate that the amount of ejected mass following AIC is largely affected by the EOS and the methods in which the transitions between regimes are handled. To show that, we use a 1D, Lagrangian, Newtonian hydrodynamics scheme and focus on an accurate treatment of the EOS and the nuclear reaction network in order to resolve the uncertainties associated with these parameters. These steps are necessary in order to simulate AIC, but they are not sufficient. In order to completely describe AIC, neutrino emission and transport, as well as GR effects, must be taken into account. These processes have a large effect on the collapse dynamics, considering the high densities and temperatures that take place during the process, and are being omitted at this point. Consequently, we do not aim to fully solve AIC in this work, but to provide the necessary first step before solving the full problem. Our solution for this case may serve as the foundation for adding more physical processes, such as the effect of deleptonization and neutrinos, GR corrections, rotation, and additional dimensions. 
Calculations were done with the VULCAN 1D hydrodynamic code \citep{Livne1993}, together with MESA routines for the EOS and for the nuclear reaction network \citep{paxton2010modules,paxton2015modules}, modified for our purposes.

\subsection{Paper Structure}

The structure of the paper is as follows: In section \ref{sec:methods}, we describe the hydrodynamic scheme we use for the simulations. Section \ref{sec:previous_studies} describes our attempts to reproduce the results of \citet{Fryer1999}. In section \ref{sec:otherprofiles}, we describe the results of simulations with our own initial profiles. In section \ref{sec:model}, we propose a simple model to describe the outcome of the collapse for initial profiles with isentropic cores. In appendix \ref{app:numerics}, we provide additional numerical details regarding the scheme, while appendix \ref{app:analytic} contains a comparison with an analytic solution of the collapse for an ideal gas polytrope \citep{Yahil1983}.

\section{Methods}
\label{sec:methods}

 %We describe the EOS used by the scheme and briefly summarize the treatment of other input physics, elaborated in appendix \ref{app:numerics}. 
We use the 1D Lagrangian hydrodynamics code VULCAN 1D \citep{Livne1993} along with our modifications. One such modification is replacing the energy with the entropy as an EOS variable. The reason is that the electrons, which contribute most of the pressure, are highly degenerate, and therefore small deviations in the energy can lead to large fluctuations in the temperature. For more details, see Appendix~\ref{app:entropy}.

The EOS consists of a few terms: electron-positron plasma, radiation, nuclei, nuclear level excitations, and Coulomb corrections, where full ionization is assumed at all times. The electrons and positrons are treated with the Timmes EOS \citep{timmes1999accuracy} at all regimes. At low densities ($\rho \ll 10^{11}\text{g}\:\text{cm}^{-3}$), nuclei are treated as an ideal gas mixture, described by the density, $\rho$, the specific entropy, $s$, and the mass fraction of each isotope, ${X_i}$. We use the MESA routines \citep{paxton2010modules} for the EOS of the ions at low densities and for the Coulomb corrections, with some modifications, to accurately describe the entropy (for more details, see \citet{kushnir2019}). Coulomb corrections are based on \citet{chabrier1998equation}. It is important to include all isotopes with nonnegligible mass fractions at the densities and temperatures where the MESA EOS is active. We have found that a list of 183 isotopes is sufficient (for more details about the isotope selection process, see \ref{app:reactions}). 

The ions are assumed to be in NSE at high temperatures and low densities, with the composition being determined by the density, $\rho$, entropy, $s$ and electron fraction, $Y_e$. At this regime, and at high densities, a tabulated EOS, which describes matter in NSE and at nuclear densities, is used (see appendix~\ref{app:reactions} for more details).

As densities approach $\sim10^{11}\,\text{g}\,\text{cm}^{-3}$, nuclear interactions become important and the ions can no longer be treated as an ideal gas. For these regimes, we use the tables and routines provided in stellarcollapse.org/microphysics \citep{o2010new,da2017new}. These tables use the nuclear EOS \citep{lattimer1991generalized,Shen1998,shen2011relativistic,shen2011new} and, as in the NSE case, receive the triplet $(\rho,s,Y_e)$ and return the rest of the variables. As the density approaches nuclear densities ($\sim10^{14}\,\text{g}\,\text{cm}^{-3}$), these EOSs become stiff, although there is some variability in modeling the transition to nuclear densities \citep{lattimer2012nuclear}. The routines provided by \citet{da2017new} allow the creation of an EOS table for matter in NSE and a table for the nuclear EOS using the single nucleus approximation \citep[SNA;][]{lattimer1991generalized} and their merger in a thermodynamically consistent manner.
This is done by summing the free energy density $ F $ of both EOSs at a crossover region:
\begin{equation}\label{key}
	F = \xi(\rho)F_\text{SNA}+(1-\xi(\rho))F_\text{NSE},
\end{equation}
where $ \xi(\rho) $ is the volume fraction of the SNA EOS, given by
\begin{equation}\label{volume_frac}
\xi(\rho)=\frac{1}{2}\left\{1+\tanh\left[ \frac{\log_{10}\left(\rho\right) - \log_{10}\left(\rho_t\right)}{\rho_\delta} \right]\right\},
\end{equation}
the transitional density is $ \rho_t=1.7\times 10^{11}\text{g}\:\text{cm}^{-3}$ and the dimensionless thickness is $ \rho_\delta=0.33 $.
We have made some modifications in these routines to fit smoothly to the modified MESA EOS, such as the Coulomb correction term. In what follows, we add "/NSE" to the name of the high-density EOS in order to describe the resulting EOS. When the EOSs of \citet{o2010new} are used, the entire table is provided in advance with no editing option, so the transition between the MESA and the tabulated nuclear EOS is not guaranteed to be smooth.

\begin{figure}
	\centering
	\includegraphics[width=\columnwidth]{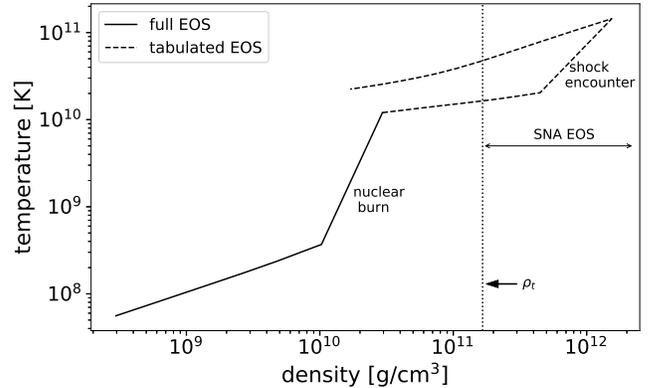}
	\caption{Density-temperature trajectory of a typical mass element during AIC (taken from the profile described in Section \ref{sec:example}). At the early stages of the collapse, matter travels along a low-density and low-temperature isentrope, described by the full EOS (solid line). When the temperature and density are high enough, nuclear burning takes place, quickly increasing the entropy and changing the composition to NSE. At this stage, the tabulated EOS is being used (dashed line). The cell then crosses the crossover region, centered at $\rho_t=1.7\times 10^{11}\,\text{g}\,\text{cm}^{-3}$ (vertical dotted line). Later on, the collapse halts by a shock wave, and the mass expands.}
	\label{fig:cell_traj}
\end{figure}

Figure~\ref{fig:cell_traj} shows the density-temperature trajectory of a typical mass element during AIC (taken from the profile described in Section \ref{sec:example}). At the early stages of the collapse, matter travels along a low-density and low-temperature isentrope, described by the EOS where the composition is evolved using a nuclear reaction network (full EOS; solid line). When the temperature and density are high enough, nuclear burning takes place, quickly increasing the entropy and changing the composition to NSE. At this stage, the tabulated EOS is being used (dashed line). The cell then crosses the crossover region, centered at $\rho_t=1.7\times 10^{11}\,\text{g}\,\text{cm}^{-3}$ (vertical dotted line). Later on, the collapse halts by a shock wave, and the element expands. It is evident from the figure that a large span of densities and temperatures and various transitions between different types of EOS are required to describe AIC.

At the low-temperature and low-density regimes, nuclear burning takes place and is implemented using the MESA routines \citep{paxton2015modules}, with raw reaction rates taken from the JINA reaclib database \citep{cyburt2010jina}. We keep $Y_e$ constant in each cell during the simulations, and we ignore neutrino emission, except in Section \ref{sec:neutrinos}. Neutrino physics of Section \ref{sec:neutrinos} is based on the deleptonization scheme of \citet{liebendorfer2005simple} and the thermal neutrino creation of \citet{itoh1996neutrino}. For more details, see appendix \ref{app:reactions} and \ref{app:neut}.

\section{Reproducing previous results}
\label{sec:previous_studies}

In order to test our scheme, we compared our results with the analytic solution of \citet{Yahil1983}, which describes the collapse of a star with a polytrope EOS. This solution is quite relevant, since a complicated EOS can be approximated as polytropes under some conditions (mainly due to the degeneracy of the electrons).
The simulation results deviate by less than $6\%$ from the analytic solution over a range of $10$ orders of magnitude. More details regarding the comparison to the analytic solution are given in appendix~\ref{app:analytic}. 

We next tried to reproduce the results of \citet{Fryer1999} for their initial profile (taken from \citet{woosley1992collapse}). This initial profile is taken in the midst of the collapse, where all parts of the star have already begun falling toward the center. The inner part of the star, up to an enclosed mass of $m=0.54M_{\text{tot}}$, where $M_{\text{tot}}$ is the total mass of the profile, is in NSE and has gone through some deleptonization, with $Y_e$ starting at $\sim0.4$ at the center and linearly increasing with the mass until it reaches $0.5$ at the edge of the inner part. For $m>0.54M_{\text{tot}}$, the star has not gone through nuclear burning or deleptonization. There, the composition is divided equally between carbon and oxygen (CO) until $m=0.82M_{\text{tot}}$, and the outer layers of the star are composed of helium. The temperature at the boundary of the NSE and non-NSE regions sharply drops from $\sim9\times10^9\,\text{K}$ to $\sim2\times10^7\,\text{K}$, since the nuclear reactions that had taken place at the inner parts of the star contribute energy to these parts and increase their temperature. \citet{Fryer1999} ran this initial profile for $0.2\,\text{s}$, testing the sensitivity to several parameters, such as the EOS and neutrino physics.

We tried to reproduce a run where the EOS used for the high densities was the LS EOS, with an incompressibility of the bulk nuclear matter parameter of $K_s=180\,\text{MeV}$ \citep[LS180 of][]{o2010new}, and without any neutrino treatment, for which \citet{Fryer1999} had obtained an ejected mass of $0.2 M_{\odot}$. There were several issues we encountered while trying to reproduce these results. The original profile used in \citet{woosley1992collapse} and in \citet{Fryer1999} could not be found \footnote{We thank Chris Fryer and Stan Woosley for their effort to locate this profile.}. Nevertheless, a similar profile was kindly provided to us by Eddie Baron. This profile did not exactly match the original profile, as the central density, radius, and maximum collapse speed of the profile we were given are $4.05\times10^{10}\,\text{g}\,\text{cm}^{-3}$, $1.1\times10^8\,\text{cm}$, and $3.2\times10^8\,\text{cm}\,\text{s}^{-1}$ compared to $6.53\times10^{10}\,\text{g}\,\text{cm}^{-3}$, $0.96\times10^8\,\text{cm}$ and $3.0\times10^8\,\text{cm}\,\text{s}^{-1}$, respectively. Additionally, the innermost region of the part of the star that is not in NSE, from $m=0.55M_{\text{tot}}$ up to $m=0.58 M_{\text{tot}}$, has negative entropy according to our MESA EOS. This is due to the degeneracy of the ions, which the MESA EOS does not handle, as it assumes that the ions are composed of an ideal gas. We have not modified this region, as the negative entropy does not interrupt the execution of the EOS. For these low densities, \citet{Fryer1999} assumed that the composition is in NSE at all times (it is unclear to us how this was implemented with an initial profile starting at a certain non-NSE composition).

We ran this profile with our scheme, using the LS EOS for the high-density regime and NSE for high temperatures and low densities, with nuclear burning taking place, and no neutrinos. The entropy in the region that is initially negative quickly rises to positive values owing to nuclear burning and small shock waves. The obtained ejected mass in our calculations is $\simeq0.173 M_{\odot}$, with this number converging for a resolution of 400 cells or higher, with the cells divided such that the resolution around the mass cut (the Lagrangian mass coordinate separating the star and the ejecta) is increased. This is within $~15\%$ of the results of \citet{Fryer1999}, despite all the issues we encountered.

We reran this simulation with LS220 and LS375 EOSs from \cite{o2010new}. While the results for the LS220 case were similar to the LS180 case, the LS375 case resulted in a slightly lower ejecta mass, $\simeq0.16 M_{\odot}$. We also reran with the LS220/NSE EOS and we obtained an ejecta mass of $\simeq0.095 M_{\odot}$, about half the amount obtained with the EOS of \cite{o2010new}. These results demonstrate the large impact of the low density regime of the EOS on the ejected mass (compared with the low impact of the high-density regimes), and the need to describe it accurately.

\section{A parameter study}
\label{sec:otherprofiles}

The initial profile of a collapsing WD has many parameters that can affect the ejected mass, such as the mass, composition, temperature profile, and electron fraction. We do not aim here to determine what is the initial profile, but instead we parameterize a few simple initial profiles, similar to the profile from the previous section, in order to provide an estimate for the range of possible ejected mass during AIC.

Our initial profiles consist of a Chandrasekhar-mass star with an isentropic core in a hydrostatic equilibrium with most of its mass having an adiabatic index below $4/3$. While the entropy profile from the previous section increases from $\approx0.8\,k_B/$baryon at the center up to $\approx1.5\, k_B/$baryon at the edge of the core, we choose to keep the core entropy constant for simplicity. It is assumed that the inner parts of the star have gone through nuclear burning and some deleptonization and are in NSE, with the electron fraction, $Y_e$, starting at some value $Y_e(0)$ at the center and rising linearly with the mass until it is $0.5$ at about $m=0.65M_{\text{tot}}$. Constructing the profile is done inside-out, where the density, pressure, temperature, and other thermodynamic quantities are determined from hydrostatic equilibrium and the predetermined entropy and electron fraction. As the distance from the center increases, the temperature decreases until it crosses a threshold $T_{\text{thr}}=9\times10^9\,\text{K}$, after which it is assumed that nuclear burning is negligible. In this region, the temperature drops to $2\times10^8\,\text{K}$ and the mass fraction is equally divided between carbon and oxygen. The entropy of the outer parts also drops by a constant factor, determined by the ratio between the entropy before the temperature drops and afterward, taken with the density at the transition. The initial entropy is chosen such that the total mass of the star is the Chandrasekhar mass.

The star is driven to collapse by giving its layers initial infall velocities or reducing the pressure in the center. The total kinetic energy from the infall velocities is orders of magnitude smaller than the gravitational energy, and from the kinetic energy at bounce. Since the adiabatic index is lower than $4/3$, the collapse continues until the EOS stiffens. An external pressure is applied to the star to maintain the hydrostatic condition at the beginning of the run and is kept constant until bounce. After bounce and expansion, as the pressure of the ejecta becomes smaller as it expands, the external pressure is modified so that the expansion will continue smoothly. We verified that the application of the external pressure has no effect on the ejected mass. 

\subsection{A specific example}
\label{sec:example}

\begin{figure*}
  \centering
  \subfloat[]{\includegraphics[width=0.48\textwidth]{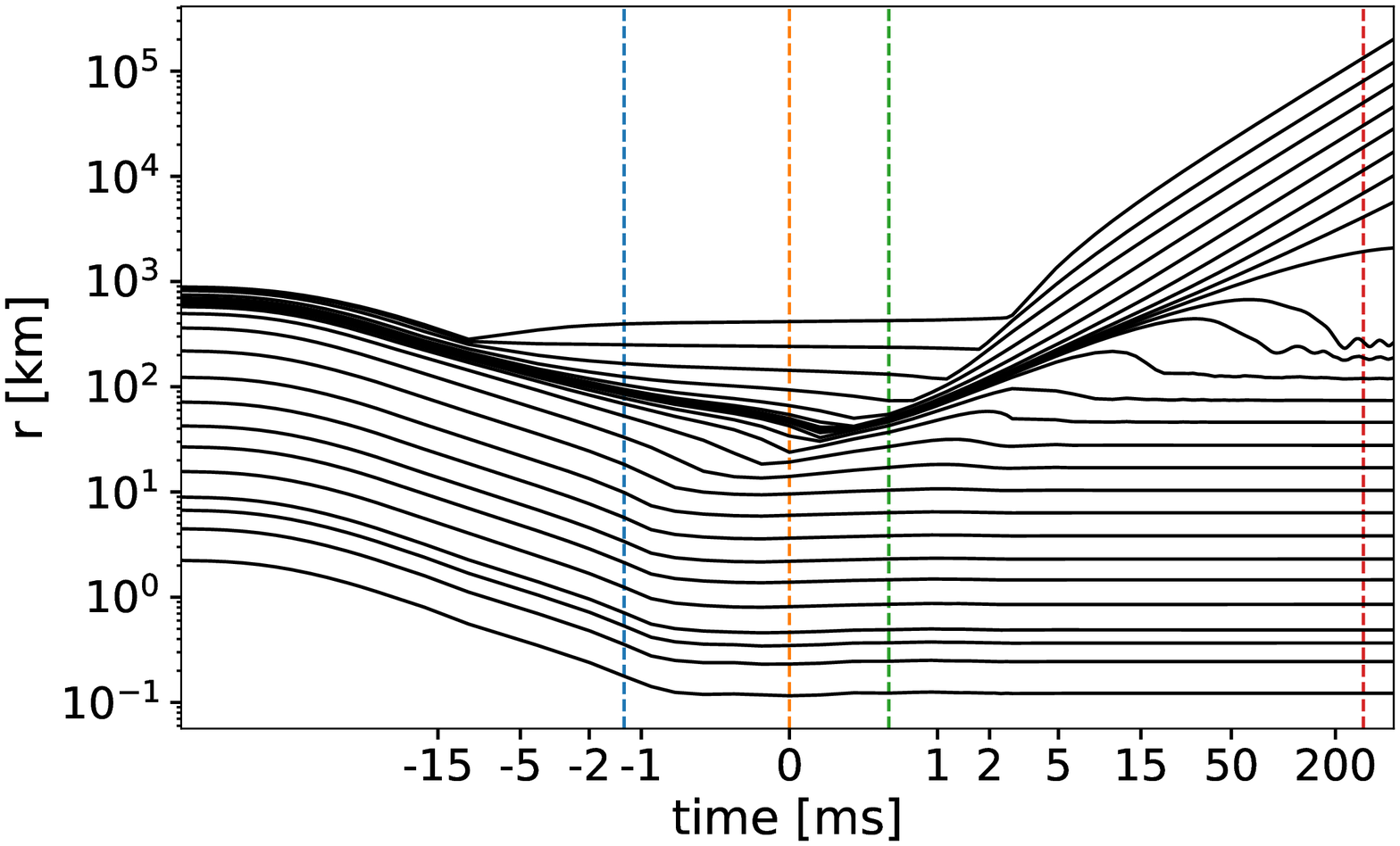}\label{fig:traj}}
  \hfill
  \subfloat[]{\includegraphics[width=0.48\textwidth]{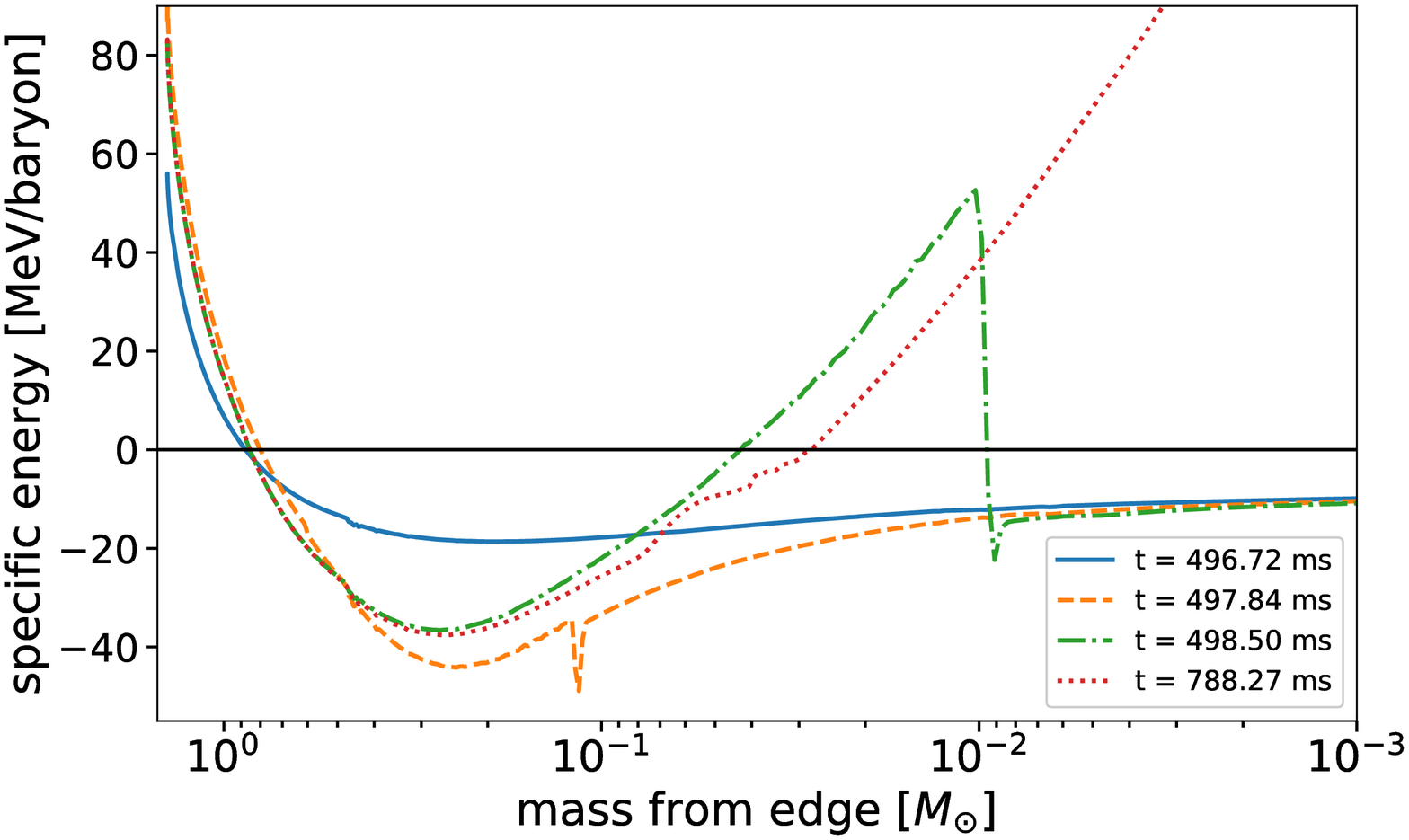}\label{fig:energies}}
  \caption{Collapse of a custom profile of a Chandrasekhar-mass WD, made with the LS220/NSE EOS, no neutrino treatment, and a resolution of 400 cells. The initial profile has an isentropic core, with an entropy of  $1.755\,k_B/\text{baryon}$. (a) Trajectories of chosen mass elements. The x-axis scale is symmetrical logarithmic, with a linear range at $t\in[-1\,\text{ms},1\,\text{ms}]$, where time is measured relative to the time of bounce. (b) Total specific energy of mass elements at different times, plotted as a function of the mass from the edge of the star. The vertical lines in panel (a) indicate the time of the curves in panel (b). These times occur before the time of bounce (blue), at the time of bounce (orange), during the shock propagation (green), and after the shock breakout (red). The troughs seen at the time of bounce (near $0.1\,M_{\odot}$) and during the shock propagation (near $0.01\,M_{\odot}$) are the shock positions. They form since these mass elements have large negative gravitational energy and small kinetic energy, and the internal energy is still increasing. The ejected mass of the collapse is calculated by the amount of mass with positive energy, which can be seen by the intersection of the energy curve with the x-axis at late times.}
  \label{fig:traj_ener}
\end{figure*}

\begin{figure}
    \centering
    \includegraphics[width=\columnwidth]{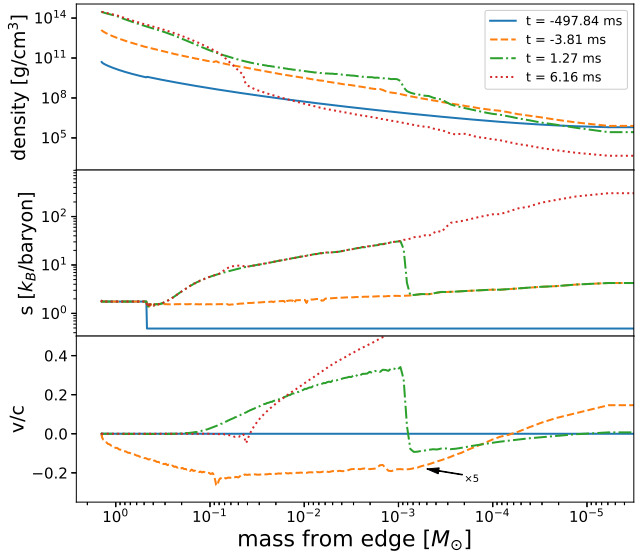}
    \caption{Simulated density (top panel), entropy (middle panel), and velocity over the speed of light (bottom panel) as a function of the the mass from the edge of the star at different epochs, for the same run as in Figure~\ref{fig:traj_ener}. Plotted are the quantities of the initial configuration (blue), before the time of bounce (orange), during the shock propagation (green), and after the shock breakout (red). The velocity during the collapse ($ t=-3.81\,[\text{ms}] $) was scaled by $ \times5 $ for convenience. Note that the times here are not the same as in Figure~\ref{fig:traj_ener}. }
    \label{fig:densities}
\end{figure}

The collapse of one specific profile is shown in Figure \ref{fig:traj_ener} and Figure \ref{fig:densities}. The nuclear EOS used in the run was the LS220/NSE EOS, created from the routines of \citet{da2017new}. The initial density and electron fraction at the center were $5\times10^{10}\,\text{g}\,\text{cm}^{-3}$ and $0.4$, respectively. The uniform initial core entropy was $1.755\,k_B/\text{baryon}$ and the resolution was 400 cells, divided with equal radial spacing. A file containing the initial structure is provided in Zenodo doi:\url{10.5281/zenodo.3740458}. We did not include deleptonization, and the whole $Y_e$ profile (as a function of mass) was kept constant throughout the collapse. Panel (a) of Figure~\ref{fig:traj_ener} shows the trajectory of chosen mass elements that represent the different trajectories, and panel (b) of Figure~\ref{fig:traj_ener} shows the total specific energy (gravitational, kinetic, and internal) of the mass elements, taken at several snapshots (before the time of bounce, at the time of bounce, during the shock propagation, and after the shock breakout). Figure \ref{fig:densities} shows the density, entropy, and velocity profiles at various times (the earliest time corresponds to the initial configuration while later epochs take place before the time of bounce, during the shock propagation and after the shock breakout). As the star collapses, the density increases in most of the star (the reason for the decrease at the edge is discussed below). After bounce, the densities reach nuclear densities of $\sim10^{14}\,\text{g}\,\text{cm}^{-3},$ and the formation of the shock wave is clearly visible. The ejected mass of this run is $2.84\times10^{-2}M_{\odot}$, which can be read from the point where the energy curve intersects with the x-axis, showing the amount of mass with positive energy. The behavior of the simulations from the previous section is qualitatively similar.

The typical energies of the ejecta are tens of $\text{MeV/baryon}$, with the outermost shells rising above $100\,\text{MeV/baryon}$. These energies correspond to mildly relativistic velocities of about $0.15c-0.3c$ for most of the ejected mass. The ejecta is composed mainly of $^{56}\text{Ni}$, since it is the most abundant isotope of matter in NSE at low temperatures and $Y_e=0.5.$ As seen in the latest epoch in Figure \ref{fig:densities}, the ejecta density profile is exponential with respect to the radius (and to the velocity, since the velocity profile is homologous) to a very good approximation.

Some mass elements started moving outward before bounce, seen at $t\simeq-10\,\text{ms}$ in Figure \ref{fig:traj} and from the decrease in the density near the edge of the star at $t\simeq-4\,\text{ms}$ in Figure \ref{fig:densities}. This is due to the energy gained by nuclear burning, causing these elements to reverse their motion before bounce. After bounce, they are quickly caught up by the emerging shock, which moves at much higher velocities (note that the energy gained by the difference of the gravitational binding energy is tens of $\text{MeV/baryon}$, an order of magnitude higher than the typical nuclear binding energy of $\sim\text{MeV/baryon}$).

\begin{figure}
    \centering
    \includegraphics[width=\columnwidth]{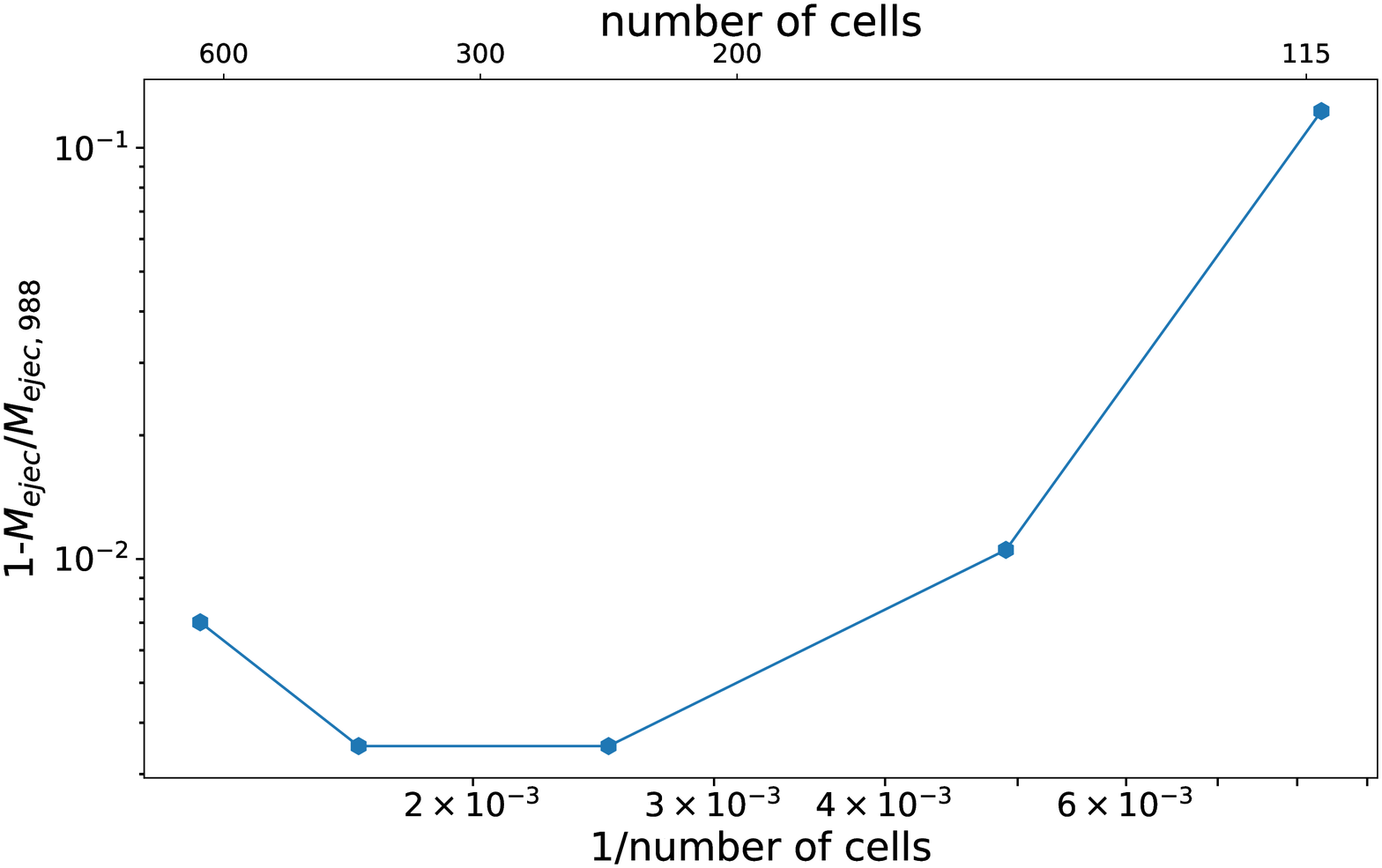}
    \caption{Deviation of the ejected mass from the highest-resolution run with 988 cells (with ejecta mass of $0.0285M_{\odot}$), as a function of the resolution. The parameters of the runs (except the number of cells) are the same as in Figure \ref{fig:traj_ener}. For resolutions higher than 200 cells, the error is within $1\%$.}
    \label{fig:convergence}
\end{figure}

In order to check convergence, we examined additional scenarios, where resolution was altered while the rest of the parameters remained the same. The resolution increased up to 988 cells that were divided in the same manner. Figure \ref{fig:convergence} shows the convergence of the run, by plotting, as a function of the resolution, the divergence of the ejected mass from that of the run with the 988 cells (where the ejected mass was $0.0285M_{\odot}$). For resolutions higher than 200 cells, the the deviation of the calculated ejected mass from that of the highest resolution does not exceed $3\times10^{-4}M_{\odot}$, reflecting a $1\%$ error.

%\begin{figure}
%	\centering
%\begin{subfigure}[b]{0.5\textwidth}
%	\centering
%	\includegraphics[width=.9\textwidth]{figures/trajv2.eps}
%	\caption{aaa}
%	\label{fig:traj}
%\end{subfigure}%
%\begin{subfigure}[b]{0.5\textwidth}
%	\centering
%	\includegraphics[width=.9\textwidth]{figures/energiesv2.eps}
%	\caption{bbb}
%	\label{fig:energies}
%\end{subfigure}
%	\caption{Collapse of a custom profile of a Chandrasekhar mass WD, made with the LS220/NSE EOS, no neutrino treatment and a resolution of 400 cells. The initial profile is isentropic, with an entropy of  $1.72\,k_B/\text{baryon}$. (a) Trajectories of chosen mass elements. The x-axis scale is symmetrical logarithmic, with a linear range at $t\in[-1\,\text{ms},1\,\text{ms}]$, where time is measured relative to the time of bounce. (b) Total specific energy of mass elements at different times, plotted as a function of the mass from the edge of the star. The vertical lines in panel (a) indicate the time of the curves in panel (b). These times occur before the time of bounce (blue), at the time of bounce (orange), during the shock propagation (green) and after the shock breakout (red). The ejected mass of the collapse is calculated by the amount of mass with positive energy, which can be seen by the intersection of the energy curve with the x-axis at late times.}
%	\label{fig:traj_ener}
%\end{figure}

\subsection{Varying the EOS and initial conditions}
\label{sec:varying}

\begin{figure}
    \centering
    \includegraphics[width=\columnwidth]{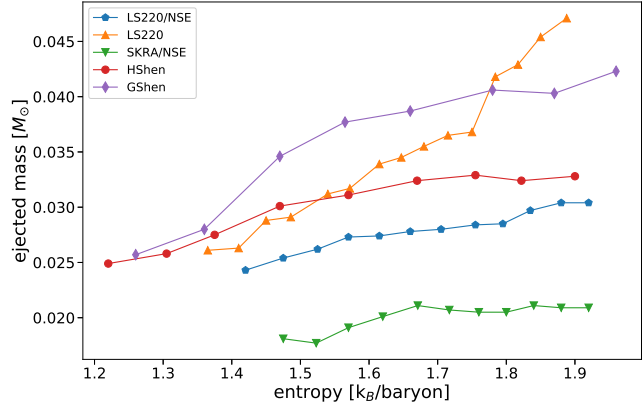}
%    \caption{Ejected mass for various EOS and initial entropy values (solid lines). Also shown are runs with no burning (purple) and the model estimates for these runs (brown dashed line). All profiles have the Chandrasekhar mass and a central density of $5\times10^{10}\,\text{g}\,\text{cm}^{-3}$. The ejected mass in all runs is a few $\times10^{-2}M_{\odot}$. The model is not valid in the low ejecta region, where another shock is created from infalling matter that further heats the outer layers of the star.}
	 \caption{Ejected mass for various EOSs and initial entropy values (solid lines). All profiles have the Chandrasekhar mass and a central density of $5\times10^{10}\,\text{g}\,\text{cm}^{-3}$. The ejected mass in all runs is a few $\times10^{-2}M_{\odot}$.}
    \label{fig:ejectavseos}
\end{figure}

In order to assess the effects of the EOS and initial conditions, we calculated the ejected mass for several nuclear EOSs and initial uniform core entropy values. The ejected mass in these runs is provided in Figure \ref{fig:ejectavseos}. The LS220/NSE and SKRA/NSE EOSs are created from the routines of \citet{da2017new} with different nuclear parameters \citep{lattimer1991generalized,rashdan2000skyrme}, and the EOSs of HShen \citep{Shen1998} and GShen \citep[][with the NL3 parameterization]{shen2011new} are taken from \citet{o2010new}. The LS180 EOS used in the previous section is not part of the list of EOSs used here, because its parameters, and specifically the incompressibility of the bulk nuclear matter being $K_s=180\,\text{MeV}$, are currently not favored \citep{lattimer2012nuclear}. The LS220 of \citet{o2010new} is also shown, for comparison, and has larger ejecta masses than the LS220/NSE, as in Section \ref{sec:previous_studies}.
%Also shown are runs where the entire star is in NSE and there is no nuclear reaction network, along with a model estimate, which is discussed in section \ref{sec:model}.

The ejected masses in Figure \ref{fig:ejectavseos} are plotted as a function of the uniform initial core entropy, all starting with a central density of $5\times10^{10}\,\text{g}\text{cm}^{-3}$. Note that for each value of the initial core entropy there is a value for the central electron fraction $Y_{e,c}$ that ensures that the mass of the star remains the Chandrasekhar mass. Since the pressure rises with the entropy and with $Y_{e,c}$, the electron fraction should be decreased for a larger initial core entropy in order to keep the mass of the star unchanged. As in the previous section, deleptonization is not included. The ejected mass in all runs is a few$\times10^{-2}M_{\odot}$. The EOS and initial core entropy alter the ejected mass by a factor of a few but do not change it by orders of magnitude. The EOSs with the low-density treatment of \citet{da2017new} all result in lower ejected mass, showing the importance of treating this region correctly. \cite{yasin2020equation} argued that the nucleon effective mass in the high-density part of the EOS,  $\rho\simeq3\times10^{14}\,\text{g}\,\text{cm}^{-3}$, plays an important role in collapsing stars, with larger effective masses leading to a faster shock evolution and explosion. This might also be important when determining the ejecta mass in AIC studies \citep[note that neutrinos, which are neglected here, play a large role in the work of][]{yasin2020equation}. Comparing EOSs with the same low-density treatment, we find that the large effective mass of the LS220/NSE EOS has a higher ejected mass than the low nucleon effective mass EOS of SKRA/NSE. Comparing the LS220 to the HShen EOS and the GShen EOS, which also have a low effective mass, seem to agree with this only at high specific entropies. 

The initial profile also has a considerable effect, as the results of this section are smaller by a factor of a few than the results in section \ref{sec:previous_studies}, where the ejected mass is $\simeq0.095M_{\odot}$ for the LS220/NSE EOS. To explain this discrepancy, we attempted to build a hydrostatic initial profile with the same entropy structure as in the initial profile in Section~\ref{sec:previous_studies} (with non-isentropic core), where the negative entropy region was replaced by a low entropy value. The total mass in the initial profile we built, using the LS220/NSE EOS, was $\sim0.05M_{\odot}$ less than the Chandrasekhar mass of the original profile, with this mass deficiency affecting the density profile at the outer parts of the star. The cause for this discrepancy is the hydrostatic equilibrium requirement that does not hold for the original profile. The ejected mass of our own hydrostatic profile was about $0.025M_{\odot}$, similar to the results in Figure \ref{fig:ejectavseos} for the LS220/NSE EOS.

We have also created a Chandrasekhar mass version of the same profile, by slightly altering the entropy structure at the core. The extra mass was mainly added to the core and not to the outer layers. The ejected mass in this case is $\approx0.032M_{\odot}$, compared to $\approx0.095M_{\odot}$ from the original profile (see Section~\ref{sec:previous_studies}), despite the same progenitor mass of both profiles. This illustrates that the original profile's excess mass of $0.05M_{\odot}$ in its outer layers, which cannot be in hydrostatic equilibrium in our profiles, is the main cause of the large ejected mass of this profile.
	
Calculations without nuclear burning, where the entire star is assumed to begin in NSE, are described in Section~\ref{sec:model} and the obtained ejecta masses are shown in Figure~\ref{fig:ejectavseos_model}. In this case, much higher ejecta masses were obtained, further demonstrating the importance of accurately describing the low-density regime and the outer layers of the progenitor star.

\subsection{Effects of neutrinos}
\label{sec:neutrinos}

Neutrino creation and emission, which are not taken into account in the previous simulations, will lead to energy and entropy losses, making it more difficult for the mass shells to escape from the star. On the other hand, the neutrinos can deposit some of the energy back when interacting with the outer layers, before escaping the star. To grasp the effect of the neutrinos, we tested the case where the neutrinos only carry energy away without any additional interaction (see appendix~\ref{app:neut}). We stress that since this is not the main focus of the paper, we only use a crude approximation for the effect of neutrinos. Detailed calculations are required to fully investigate the effects of neutrinos. The addition of neutrino emission, mainly due to deleptonization, leads to a reduction in the ejected mass by roughly an order of magnitude, to $\sim10^{-3}M_{\odot}$, for the same initial profiles of Section~\ref{sec:varying}. The simulations include $\sim300$ cells, where the outer cell mass resolution is $\sim10^{-4}\,M_{\odot}$. This can be considered a lower limit for these profiles, EOSs and input physics.

%As explained in \ref{app:neut}, this is not an accurate treatment of the neutrino physics, but to get some sense of how the neutrinos affect the process.

\section{A simple estimate of the ejected mass}
\label{sec:model}
\begin{figure}
    \centering
    \includegraphics[width=\columnwidth]{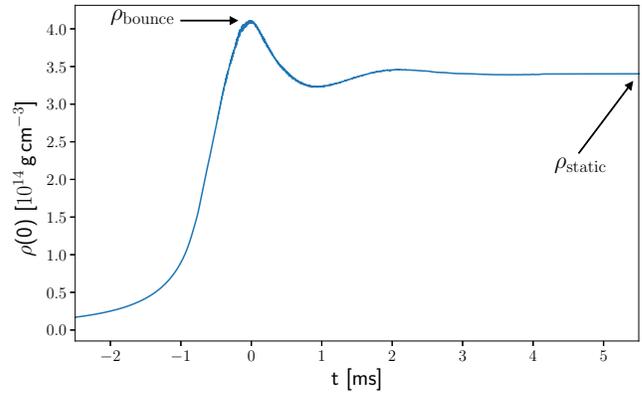}
    \caption{Central density as a function of time for a typical simulation run. The two unknown parameters of the model, the central density at bounce and the central density of the hydrostatic star, are marked.}
    \label{fig:rho_vs_t}
\end{figure}

We now construct a simple model to describe the structure of the PNS at the end of the collapse and to evaluate the ejected mass. We assume that the tabulated EOS, which describes nuclear matter for high densities and ideal gas in NSE for low densities, can be used for the entire star. We therefore assume that the matter is in NSE and that there is no burning, and we neglect neutrinos. Another assumption is, as in the previous section, that the initial profile is isentropic (throughout the entire star in this case). We assume that the flow is isentropic throughout the collapse, until the matter is shock heated by the shock that was created from the bounce of the infalling matter, which alters the entropy profile, resulting in hydrostatic equilibrium. The model requires two unknown parameters - the central density of the final hydrostatic star, $\rho_{\text{static}}$, and the (larger) central density at the time of bounce, $\rho_{\text{bounce}}$, which is used in order to estimate the shock strength. 
The two parameters, together with the hydrostatic equilibrium and isentropic flow conditions above, are sufficient to construct the PNS. The two parameters are retrieved from the numerical simulations in the previous section. Figure \ref{fig:rho_vs_t} shows the central density as a function of time for a typical run, and how we extracted the density of the hydrostatic star and at bounce that we used in the model. The ejected mass can be calculated from the difference in mass between the initial WD and the PNS. 

To begin constructing a hydrostatic star, the central density, $\rho_{\text{static}}$, and entropy (or any other two thermodynamic variables, given an EOS) should be predetermined. The latter is determined from the initial entropy, while the former, as mentioned above, is extracted from the simulation result. The final central density can be approximately estimated as the nuclear density, but the ejected mass was found to be quite sensitive to this value, changing by $10-30\%$ as a result of a $5\%$ change in the density. The density-pressure profile is then determined from the hydrostatic equilibrium,
\begin{equation}
   \frac{\partial p}{\partial r} = -\frac{Gm\rho}{r^2},
\end{equation} 
and the shock heating. To estimate the shock heating, we assume that the infalling matter bounced and immediately stopped; hence, the strength of the shock will be determined out of the infall velocity, $u$, through the Hugoniot conditions,
\begin{align}
   &u^2 = \left(p_1-p_0\right)\left(V_0-V_1\right)\\
   &\epsilon_1 - \epsilon_0 = \frac{1}{2}\left(p_1+p_0\right)\left(V_0-V_1\right),
\end{align} 
where $V=1/\rho$ and the 0 and 1 subscripts refer to the upstream and downstream, respectively. Once the infall velocity is known, it is possible to retrieve all the variables using the conditions mentioned above. 

To estimate the infall velocity, we make use of the self-similar solution discussed in Appendix~\ref{app:analytic}. The velocity is given by
\begin{equation}
    u=\kappa^{\frac{1}{2}}G^{\frac{1-\gamma}{2}}\left (t_c-t_{\text{bounce}}  \right)^{1-\gamma}U(X),
\end{equation}
where $\kappa=p/\rho^\gamma$; $X$ and $U$ are the known self-similar coordinate and velocity, respectively; and $t_c-t_{\text{bounce}}$ is the time from bounce (for the self-similar solution, the density diverges at $t_c$).
The adiabatic index $\gamma$ and the constant $\kappa$ are determined from the EOS at the given initial entropy and an averaged value of $Y_e$. The quantity $t_c-t_{\text{bounce}}$ is related to the second unknown parameter - the central density at bounce -- by using the (known) self-similar density:
\begin{equation}
\label{eq:ss dens}
    D(0)=G\left(t_c -t_{\text{bounce}}\right)^2\rho_{\text{bounce}}.
\end{equation}
In such a way, the velocity can be obtained given the radius, $r$, and the PNS can be constructed.

We note that the quantity $t_c-t_{\text{bounce}}$ can be roughly estimated by comparing the final central density (the first parameter) to the self-similar density using the same relation, Equation~\eqref{eq:ss dens}, but this gives an overestimate of $t_c-t$, since the density of the end state is not as high as it was at the time of the bounce and shock formation. This will result in a weaker shock and lower temperatures and pressures, which can support less mass, leading to a higher ejected mass. The estimation of $t_c-t_{\text{bounce}}$ from $\rho_{\text{bounce}}$ is not entirely optimal but gives the correct ejected mass to within $15\%$ for the LS220 and the LS220/NSE EOSs, where in some cases the difference is less than $5\%$. When the initial entropy decreases and the ejected mass becomes low ($\lesssim 0.035M_{\odot}$), some of the matter being ejected by the shock falls back into the star. When it meets the star, another shock wave is created, which further increases the entropy of this matter. This effect is not captured in our model, so the estimates for the runs with low initial entropy give a lower mass compared to the simulations.

\begin{figure}
	\centering
	\includegraphics[width=\columnwidth]{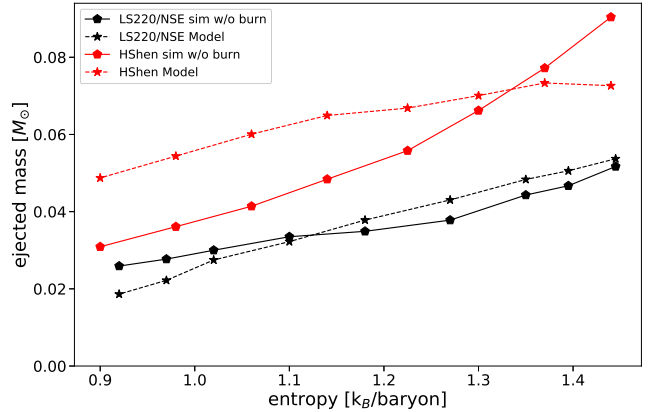}
	\caption{Simulation and model ejected mass for the LS220/NSE (black) and the HShen (red) EOS, as a function of the initial entropy. The simulation results are shown in pentagons and solid lines, and the model results are shown in stars and dashed lines. All profiles have the Chandrasekhar mass and a central density of $5\times10^{10}\,\text{g}\,\text{cm}^{-3}$. The model and the simulation results of the LS EOS fit quite well, while for the HShen EOS the differences are somewhat larger.}
	%	 The model is not valid in the low ejecta region, where another shock is created from infalling matter that further heats the outer layers of the star. 
	\label{fig:ejectavseos_model}
\end{figure}

Figure~\ref{fig:ejectavseos_model} presents the simulation results and the model estimations for profiles with the LS220/NSE and HShen EOSs and an initial central density of $5\times10^{10}\,\text{g}\,\text{cm}^{-3}$. The model results for the HShen EOS are not as accurate as for the LS220 EOS but provide the correct results to within $50\%$. 
%Amir: is the following sentence necessary? For the SKRA/NSE EOS, the estimated mass of the star from the model was high as the progenitor mass, so the model estimates of the ejected masses were very low for this EOS.}
%The regime of ejected mass where the second shock is significant is marked, and the model's deviations grow substantially.
Since the model calculates the ejecta from the mass difference between the two stars, it is valid only because the ejecta is not a very small fraction of the star, about $2-5\%$. For an ejecta of $5\%$ of the star, an error of $1\%$ of the mass of the PNS would result in an error of $20\%$ and will increase as the ejecta becomes a smaller fraction. If the range of ejected mass would have been an order of magnitude smaller, the accuracy of the model would have to be high enough to calculate the mass of the PNS within $<0.1\%$ in order to have reasonable results. 

\begin{figure}
	\centering
	\includegraphics[width=\columnwidth]{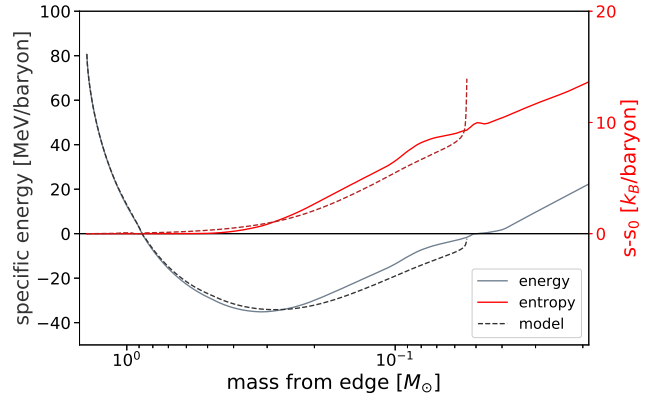}
	\caption{Results of the simulation (solid lines) and the model (dashed lines) for the energy and entropy as a function of the mass from the edge of the star. For the simulation, we used the LS220/NSE EOS, initial specific entropy of $1.45 \,k_B/\text{baryon}$ and $Y_e(0)=0.40$. Plotted are the total specific energy (gray) and the change in the initial entropy (red). The ejected mass of the simulation is the mass where the energy becomes positive, and for the model it is simply where the curve ends. The model predicts the ejected mass to an accuracy of $\sim4\%$. The deviation of the entropy and energy between the simulation and model is less than $20\%$.}
	\label{fig:model}
\end{figure}

Figure \ref{fig:model} compares the energy and entropy profiles of the simulation to the ones derived with the simple model for a run with the LS220/NSE EOS, starting with an initial central density of $5\times10^{10}\,\text{g}\,\text{cm}^{-3}$, uniform entropy of $1.45 \,k_B/\text{baryon}$ and $Y_e(0)=0.4$. The mass with positive energy is being ejected. The entropy change shows how well the shock strength is described by the model, since it is independent of the hydrostatic equilibrium constraint, unlike the other parameters, such as density and pressure. The ejected mass of this run is $0.0517\,M_{\odot}$, while the model gives a mass of $0.0537M_{\odot}$, a difference of $\sim4\%$.
% To indicate the differences, the model is represented by a curve when fitting $t_c-t_{\text{bounce}}$ with the density at bounce, and a curve when the final central density is used in its estimation. The second curve, corresponding to a weaker shock, under-estimates the mass of the star and over-estimates the ejecta by almost $50\%$, while the first curve fits much better to the simulation and correctly estimates the ejected mass. 
%This is to be expected, since by fitting the parameter $t_c-t_{\text{bounce}}$ it is possible to get any desired ejected mass. However, 

The deviation in the density and entropy profiles between the model and the simulation is less than $2\%$ in density and less than $20\%$ in entropy, except for the model's last part, $10^{-3}M_{\odot}$ from the edge, where the entropy quickly rises.

\section{Conclusions}

In this work, we implemented a numerical scheme to calculate the amount of ejected mass following AIC. We assumed a spherical collapse and neglected neutrinos and GR corrections, while accurately treating the EOS and nuclear burning. We found the amount of ejected mass to be a few $\times10^{-2}M_{\odot}$ for a large range of initial conditions and EOSs, and, as expected, to always move at mildly relativistic velocities.
The low-density regime of the EOS was found to have a nonnegligible impact on the EOS, as seen in figures \ref{fig:ejectavseos} and \ref{fig:ejectavseos_model}, which show that all regimes in the EOS should be accurately treated. We have suggested a simple model that estimates the ejected mass at the end of the AIC, for the LS EOS, to within $15\%$ given two free parameters, while describing the PNS structure to reasonable accuracy ($<2\%$ in density and $<20\%$ in entropy). 

Our numerical scheme can serve as a basis for future studies, where more physical aspects can be taken into account and integrated into the scheme. Such processes were already studied in the context of AIC, and include GR \citep{Fryer1999}, neutrino physics \citep{woosley1992collapse,Fryer1999,Dessart2006} and rotation \citep{Dessart2006}. \cite{marek2006exploring} developed a method to include GR corrected potential for a Newtonian hydrodynamic code, which could be integrated in an AIC simulation as well. We believe that our work is essential for any study that aims to accurately describe the AIC process.

%The initial profiles used were of our own creation, and do not aim to represent the undetermined WD progenitor.
\citet{Waxman2017} suggested that the persistent radio source, associated with FRB 121102, can be explained by the propagation of a spherical shell of plasma into a surrounding medium. The properties of the persistent source imply that the mass of the shell is $\sim 10^{-5}M_{\odot}$ moving at mildly relativistic velocities, for a duration of $t<10^{2.5}\,\text{yr}$. This suggests that the source was created by a weak stellar explosion, such as AIC, since the mass of the shell is much smaller than the ejecta mass from a typical supernova. The obtained ejected mass of $\sim10^{-2}M_{\odot}$ is larger than the estimated ejected mass shell of the FRB 121102 persistent source ($\sim10^{-5}M_{\odot}$). This inconsistency could be a result of the neglected physical processes, such as neutrino physics, GR corrections and rotation, or nonrepresentative initial conditions. A preliminary neutrino emission calculation did reduce the ejecta mass by an order of magnitude, but not down to $\sim10^{-5}M_{\odot}$. Future work could determine more reliably the connection between the persistent source of FRB 121102 and the AIC process.

\section*{Acknowledgements}
We thank Eddie Baron, Boaz Katz, Eran Ofek, and Eli Waxman for useful discussions. D.K. is supported by the Israel Atomic Energy Commission, the Council for Higher Education, the Pazi Foundation, and by a research grant from The Abramson Family Center for Young Scientists.

%%%%%%%%%%%%%%%%%%%%%%%%%%%%%%%%%%%%%%%%%%%%%%%%%%

%%%%%%%%%%%%%%%%%%%% REFERENCES %%%%%%%%%%%%%%%%%%

% The best way to enter references is to use BibTeX:
\bibliographystyle{aasjournal}
\bibliography{bibliography} % if your bibtex file is called example.bib

\begin{thebibliography}{}
\expandafter\ifx\csname natexlab\endcsname\relax\def\natexlab#1{#1}\fi
\providecommand{\url}[1]{\href{#1}{#1}}
\providecommand{\dodoi}[1]{doi:~\href{http://doi.org/#1}{\nolinkurl{#1}}}
\providecommand{\doeprint}[1]{\href{http://ascl.net/#1}{\nolinkurl{http://ascl.net/#1}}}
\providecommand{\doarXiv}[1]{\href{https://arxiv.org/abs/#1}{\nolinkurl{https://arxiv.org/abs/#1}}}

\bibitem[{Baron {et~al.}(1985)Baron, Kahana, \& Cooperstein}]{baron1985type}
Baron, E., Kahana, S., \& Cooperstein, J. 1985, PhRvL, 55, 126

\bibitem[{Bethe(1990)}]{bethe1990supernova}
Bethe, H.~A. 1990, Reviews of Modern Physics, 62, 801

\bibitem[{{Canal} {et~al.}(1980){Canal}, {Isern}, \& {Labay}}]{canal1980r}
{Canal}, R., {Isern}, J., \& {Labay}, J. 1980, ApJL, 241, L33,
  \dodoi{10.1086/183354}

\bibitem[{Chabrier \& Potekhin(1998)}]{chabrier1998equation}
Chabrier, G., \& Potekhin, A.~Y. 1998, Physical Review E, 58, 4941

\bibitem[{Chatterjee {et~al.}(2017)Chatterjee, Law, Wharton, Burke-Spolaor,
  Hessels, Bower, Cordes, Tendulkar, Bassa, Demorest,
  {et~al.}}]{chatterjee2017direct}
Chatterjee, S., Law, C., Wharton, R., {et~al.} 2017, Nature, 541, 58

\bibitem[{Cyburt {et~al.}(2010)Cyburt, Amthor, Ferguson, Meisel, Smith, Warren,
  Heger, Hoffman, Rauscher, Sakharuk, {et~al.}}]{cyburt2010jina}
Cyburt, R.~H., Amthor, A.~M., Ferguson, R., {et~al.} 2010, ApJS, 189, 240

\bibitem[{Darbha {et~al.}(2010)Darbha, Metzger, Quataert, Kasen, Nugent, \&
  Thomas}]{darbha2010nickel}
Darbha, S., Metzger, B., Quataert, E., {et~al.} 2010, Monthly Notices of the
  Royal Astronomical Society, 409, 846

\bibitem[{Dessart {et~al.}(2006)Dessart, Burrows, Ott, Livne, Yoon, \&
  Langer}]{Dessart2006}
Dessart, L., Burrows, A., Ott, C.~D., {et~al.} 2006, ApJ, 644, 1063,
  \dodoi{10.1086/503626}

\bibitem[{Drout {et~al.}(2014)Drout, Chornock, Soderberg, Sanders, McKinnon,
  Rest, Foley, Milisavljevic, Margutti, Berger, {et~al.}}]{drout2014rapidly}
Drout, M.~R., Chornock, R., Soderberg, A.~M., {et~al.} 2014, ApJ, 794, 23

\bibitem[{Fryer {et~al.}(1999)Fryer, Benz, Herant, \& Colgate}]{Fryer1999}
Fryer, C., Benz, W., Herant, M., \& Colgate, S.~A. 1999, ApJ, 516, 892,
  \dodoi{10.1086/307119}

\bibitem[{{Herant} {et~al.}(1994){Herant}, {Benz}, {Hix}, {Fryer}, \&
  {Colgate}}]{herant1994}
{Herant}, M., {Benz}, W., {Hix}, W.~R., {Fryer}, C.~L., \& {Colgate}, S.~A.
  1994, \apj, 435, 339, \dodoi{10.1086/174817}

\bibitem[{Itoh {et~al.}(1996)Itoh, Hayashi, Nishikawa, \&
  Kohyama}]{itoh1996neutrino}
Itoh, N., Hayashi, H., Nishikawa, A., \& Kohyama, Y. 1996, ApJS, 102, 411

\bibitem[{Kashiyama \& Murase(2017)}]{kashiyama2017testing}
Kashiyama, K., \& Murase, K. 2017, ApJL, 839, L3

\bibitem[{Kushnir(2018)}]{kushnir2019}
Kushnir, D. 2018, MNRAS, 483, 425, \dodoi{10.1093/mnras/sty3121}

\bibitem[{Lattimer(2012)}]{lattimer2012nuclear}
Lattimer, J.~M. 2012, Annual Review of Nuclear and Particle Science, 62, 485

\bibitem[{Lattimer \& Swesty(1991)}]{lattimer1991generalized}
Lattimer, J.~M., \& Swesty, F.~D. 1991, Nuclear Physics A, 535, 331

\bibitem[{Liebend{\"o}rfer(2005)}]{liebendorfer2005simple}
Liebend{\"o}rfer, M. 2005, ApJ, 633, 1042

\bibitem[{Liebend\"orfer {et~al.}(2001)Liebend\"orfer, Mezzacappa, Thielemann,
  Messer, Hix, \& Bruenn}]{liebendorfer2001}
Liebend\"orfer, M., Mezzacappa, A., Thielemann, F.-K., {et~al.} 2001, PhRvD,
  63, 103004, \dodoi{10.1103/PhysRevD.63.103004}

\bibitem[{Livne(1993)}]{Livne1993}
Livne, E. 1993, ApJ, 412, 634, \dodoi{10.1086/172950}

\bibitem[{Lyne {et~al.}(1996)Lyne, Manchester, \& D'Amico}]{lyne1996psr}
Lyne, A., Manchester, R., \& D'Amico, N. 1996, ApJL, 460, L41

\bibitem[{{Lyutikov} \& {Toonen}(2019)}]{lyutikov2018fbots}
{Lyutikov}, M., \& {Toonen}, S. 2019, \mnras, 487, 5618,
  \dodoi{10.1093/mnras/stz1640}

\bibitem[{Marek {et~al.}(2006)Marek, Dimmelmeier, Janka, M{\"u}ller, \&
  Buras}]{marek2006exploring}
Marek, A., Dimmelmeier, H., Janka, H.-T., M{\"u}ller, E., \& Buras, R. 2006,
  A\&A, 445, 273

\bibitem[{Margalit {et~al.}(2019)Margalit, Berger, \&
  Metzger}]{margalit2019fast}
Margalit, B., Berger, E., \& Metzger, B.~D. 2019, The Astrophysical Journal,
  886, 110

\bibitem[{Metzger {et~al.}(2009)Metzger, Piro, \& Quataert}]{Metzger2009}
Metzger, B.~D., Piro, A.~L., \& Quataert, E. 2009, MNRAS, 396, 1659,
  \dodoi{10.1111/j.1365-2966.2009.14909.x}

\bibitem[{Moriya(2019)}]{moriya2019vtc}
Moriya, T.~J. 2019, Monthly Notices of the Royal Astronomical Society, 490,
  1166

\bibitem[{Nomoto(1986)}]{nomoto1986fate}
Nomoto, K. 1986, Progress in Particle and Nuclear Physics, 17, 249

\bibitem[{O'Connor \& Ott(2010)}]{o2010new}
O'Connor, E., \& Ott, C.~D. 2010, Classical and Quantum Gravity, 27, 114103

\bibitem[{Paxton {et~al.}(2010)Paxton, Bildsten, Dotter, Herwig, Lesaffre, \&
  Timmes}]{paxton2010modules}
Paxton, B., Bildsten, L., Dotter, A., {et~al.} 2010, ApJS, 192, 3

\bibitem[{Paxton {et~al.}(2015)Paxton, Marchant, Schwab, Bauer, Bildsten,
  Cantiello, Dessart, Farmer, Hu, Langer, {et~al.}}]{paxton2015modules}
Paxton, B., Marchant, P., Schwab, J., {et~al.} 2015, ApJS, 220, 15

\bibitem[{Rashdan(2000)}]{rashdan2000skyrme}
Rashdan, M. 2000, Modern Physics Letters A, 15, 1287

\bibitem[{Schneider {et~al.}(2017)Schneider, Roberts, \& Ott}]{da2017new}
Schneider, A. d.~S., Roberts, L.~F., \& Ott, C.~D. 2017, Physical Review C, 96,
  065802

\bibitem[{Scholz {et~al.}(2016)Scholz, Spitler, Hessels, Chatterjee, Cordes,
  Kaspi, Wharton, Bassa, Bogdanov, Camilo, {et~al.}}]{scholz2016repeating}
Scholz, P., Spitler, L., Hessels, J., {et~al.} 2016, ApJ, 833, 177

\bibitem[{Shen {et~al.}(2011{\natexlab{a}})Shen, Horowitz, \&
  Teige}]{shen2011new}
Shen, G., Horowitz, C., \& Teige, S. 2011{\natexlab{a}}, Physical Review C, 83,
  035802

\bibitem[{Shen {et~al.}(1998)Shen, Toki, Oyamatsu, \& Sumiyoshi}]{Shen1998}
Shen, H., Toki, H., Oyamatsu, K., \& Sumiyoshi, K. 1998, Progress of
  Theoretical Physics, 100, 1013, \dodoi{10.1143/PTP.100.1013}

\bibitem[{Shen {et~al.}(2011{\natexlab{b}})Shen, Toki, Oyamatsu, \&
  Sumiyoshi}]{shen2011relativistic}
---. 2011{\natexlab{b}}, ApJS, 197, 20

\bibitem[{{Swesty} {et~al.}(1994){Swesty}, {Lattimer}, \&
  {Myra}}]{Swesty1994ApJ}
{Swesty}, F.~D., {Lattimer}, J.~M., \& {Myra}, E.~S. 1994, \apj, 425, 195,
  \dodoi{10.1086/173974}

\bibitem[{{Tauris, T. M.} {et~al.}(2013){Tauris, T. M.}, {Sanyal, D.}, {Yoon,
  S.-C.}, \& {Langer, N.}}]{Tauris2013}
{Tauris, T. M.}, {Sanyal, D.}, {Yoon, S.-C.}, \& {Langer, N.} 2013, A\&A, 558,
  A39, \dodoi{10.1051/0004-6361/201321662}

\bibitem[{Timmes \& Arnett(1999)}]{timmes1999accuracy}
Timmes, F., \& Arnett, D. 1999, ApJS, 125, 277

\bibitem[{Waxman(2017)}]{Waxman2017}
Waxman, E. 2017, ApJ, 842, 34, \dodoi{10.3847/1538-4357/aa713e}

\bibitem[{Woosley \& Baron(1992)}]{woosley1992collapse}
Woosley, S., \& Baron, E. 1992, ApJ, 391, 228

\bibitem[{Yahil(1983)}]{Yahil1983}
Yahil, A. 1983, ApJ, 265, 1047, \dodoi{10.1086/160746}

\bibitem[{Yasin {et~al.}(2020)Yasin, Sch{\"a}fer, Arcones, \&
  Schwenk}]{yasin2020equation}
Yasin, H., Sch{\"a}fer, S., Arcones, A., \& Schwenk, A. 2020, Physical Review
  Letters, 124, 092701

\end{thebibliography}

% Alternatively you could enter them by hand, like this:
% This method is tedious and prone to error if you have lots of references
% \begin{thebibliography}{99}
% \bibitem[\protect\citeauthoryear{Author}{2012}]{Author2012}
% Author A.~N., 2013, Journal of Improbable Astronomy, 1, 1
% \bibitem[\protect\citeauthoryear{Others}{2013}]{Others2013}
% Others S., 2012, Journal of Interesting Stuff, 17, 198
% \end{thebibliography}

%%%%%%%%%%%%%%%%%%%%%%%%%%%%%%%%%%%%%%%%%%%%%%%%%%

%%%%%%%%%%%%%%%%% APPENDICES %%%%%%%%%%%%%%%%%%%%%

\appendix
\renewcommand\thefigure{\thesection.\arabic{figure}}   
\section{Numerical details}
\label{app:numerics}

\subsection{Nuclear reaction network and NSE}
\label{app:reactions}
Nuclear reactions were calculated with the MESA routines \citep{paxton2015modules}. The forward reaction rates are taken from the JINA reaclib database \citep{cyburt2010jina}. The rates in JINA are only valid up to $10^{10}\,\text{K}$, which does not cover the temperature range in our simulations, even for matter that is not in NSE. We therefore used the rate at $10^{10}\,\text{K}$ for higher temperatures. Backward rates are calculated from detailed balance. %Detailed balance of the isotopes occurs when the matter has thermalized with respect to strong nuclear reactions, and the entropy can no longer increase from these reactions. 
We allow the EOS of each numerical cell to change from the MESA routines for ideal gas to NSE/nuclear, and vice versa. When the cell is at the MESA EOS, isotopes approach NSE owing to nuclear burning. When the temperature is higher than $5\times10^9\,\text{K}$ and the mass fraction $X_i$ of every isotope does not differ by more than 0.01 from the NSE composition, the EOS is replaced. Since the EOSs are not always identical, we kept the pressure and density of the previous EOS, while the other quantities (temperature, entropy, etc..) are set by the new EOS. The pressure and density are chosen so that the force applied by the cell and its mass remain unchanged. The NSE composition is calculated for 3335 isotopes in order to accurately describe the EOS over a wide range of $T,\rho$, and $Y_e$. In order to have a smooth transition when changing the EOS, it is required that the MESA EOS will contain all isotopes whose NSE composition at transition has nonnegligible mass fractions. The list of these isotopes was prepared in the same method as in \citet{kushnir2019}, by observing the isotopes with a molar mass fraction, $Y_i=X_i/A_i$ larger than $10^{-5}$ at the range of parameters for which the MESA EOS is relevant, $T\in [2\times10^{9},3\times10^{10}]\,\text{K}$, $\rho\in [100,10\times 10^{10}]\,\text{g}\:\text{cm}^{-3}$ and $Y_e\in[0.495,0.5]$. We have found that list NSE5 (with 179) from \citet{kushnir2019} together with four additional isotopes ($^{34}$P, $^{35}$P, $^{62}$Ga, $^{65}$Ge), is sufficient. As the temperature of a cell in NSE drops below $5\times10^9\,\text{K}$, the EOS changes to the MESA EOS, with the composition being the NSE composition at the time of the transition. The isotope list, chemical data, and reaction rates can be found on Zenodo doi: \url{10.5281/zenodo.3740458}\\

\subsection{Entropy as an EOS variable}
\label{app:entropy}
At the large densities reached during the simulations, matter is highly degenerate and the temperature is a very sensitive function of the internal energy. Since the temperature is used in all our EOS as one of the input variables, it should be accurately obtained. The dependence of the specific entropy $s$ on the temperature is not a very sensitive function, and thus the entropy, instead of the energy, is used to evolve the EOS quantities, using the second law of thermodynamics: 
\begin{equation}
    Tds = \delta Q,
\end{equation}
with $Q$ being the heat flow (per unit mass) in or out of the system. The heating can be a result of shock, neutrino emission, and nuclear burning. Note that when the composition changes owing to nuclear burning, the heat added is the sum of the energy increase due to rest mass difference and chemical potential times the composition difference: 
\begin{equation}
    \delta Q_{nuc} = de - \sum_i \mu_i dy_i,
\end{equation}
where $ \delta Q_\text{nuc} $ is the added heat, $ de $ is the added energy, and $ \mu_i $ and $ dy_i $ are the chemical potential and the change in the molar fraction of the i-th isotope, respectively.
In NSE, matter reaches thermal equilibrium with respect to nuclear reactions, and these terms cancel each other out, such that the entropy no longer changes owing to nuclear reactions. This equation, together with the equation of motion and the EOS, is solved in the hydrodynamic code. The equation of motion is
\begin{equation}
    \frac{du}{dt}=-\frac{1}{\rho}\frac{\partial p}{\partial r}-\frac{mG}{r^2},
\end{equation}
where $ r $ is the radial coordinate, $ u $ is the velocity, $ p$ is the pressure, $ \rho $ is the density, and $ m $ is the mass up to $ r $. This is done in two steps within the hydrodynamic code in a leapfrog integration scheme, where the velocities and positions are calculated at interleaved time points.

\subsection{Neutrino physics}
\label{app:neut}
Although neutrino physics has an important role at the large densities and temperatures that take place in the AIC process, it is only partly considered in this work. In order to give a lower bound to the ejected mass, we omit the interaction of the neutrinos after creation and assume they freely escape the star, reducing the energy and entropy as a result. We consider two processes - deleptonization due to electron capture and creation of thermal neutrino-anti-neutrino pairs. To account for deleptonization, we use the prescription of \citet{liebendorfer2005simple}, in which the electron fraction $Y_e$ is determined from the instantaneous density alone. This is only a crude approximation, as the prescription was calibrated from a full calculation that depends on an EOS and an initial profile that are not the ones used in our simulations. Also, this prescription is relevant only until bounce, so we let the $Y_e$ profile freeze afterward. Entropy loss is calculated by \citep{bethe1990supernova}
\begin{equation}
    Tds=-\delta y_e\left(\tfrac{1}{6}\mu _e -\left(\mu_n-\mu_p\right)\right),
\end{equation}
where $\mu_{\{e,n,p\}}$ is the chemical potential of the electrons, neutrons, and protons, respectively.

Creation and emission of thermal neutrinos are calculated from expressions $\{2.1,\,3.1,\,4.1,\,5.1,\,5.17,\,5.27,\,5.29,\,6.5\}$ of \citet{itoh1996neutrino}, taking into account all the processes that contribute to the specific energy loss rate, which are given as a function of the density, temperature, mean number of nucleons $\Bar{A}$, and mean charge $\Bar{Z}$. Since we assume that the composition does not change in this process, the energy loss is entirely converted to heat loss. This process, however, is quite negligible when deleptonization occurs.

\section{Comparison to analytic solution}
\label{app:analytic}
\setcounter{figure}{0} 
In order to test our numerical scheme, we compared it with the self-similar analytic solution of \citet{Yahil1983} for the collapse of a spherically symmetric star with a polytropic EOS:
\begin{equation}
    p=\kappa\rho^{\gamma},
\end{equation} 
for the adiabatic index $\nicefrac{6}{5}\leq \gamma \leq \nicefrac{4}{3}$, where $ p$ is the pressure, $ \rho $ is the density, and $ \kappa $ is a constant in space and time. The solution uses the self-similar dimensionless coordinate 
\begin{equation}
    % X\propto  \left (t_c-t  \right)^{\gamma-2}r,
    X = r/r_0(t),\quad r_0(t)=\kappa^{\frac{1}{2}}G^{\frac{1-\gamma}{2}}\left (t_c-t  \right)^{2-\gamma}.
\end{equation} 
The interval from the catastrophe time $t_c-t$ determines the scaling of the physical variables (coordinate, velocity, etc.) to the self-similar ones. This interval is unknown during the simulation and is computed by best fitting to the analytic solution. Figure \ref{fig:analitic} shows the velocity and density during the collapse of a star with an ideal gas EOS with $\gamma=1.3$, normalized to the self-similar variables, done with a resolution of 2000 cells, initially divided with equal radial spacing. The self-similar solution is plotted on top for comparison. The agreement is within $6\%$ over a range of 10 orders of magnitude. A convergence test we have performed shows that the deviation between the simulation and the analytic solution is (roughly) inversely proportional to the resolution.

\begin{figure}
    \centering
    \includegraphics[width=.7\columnwidth]{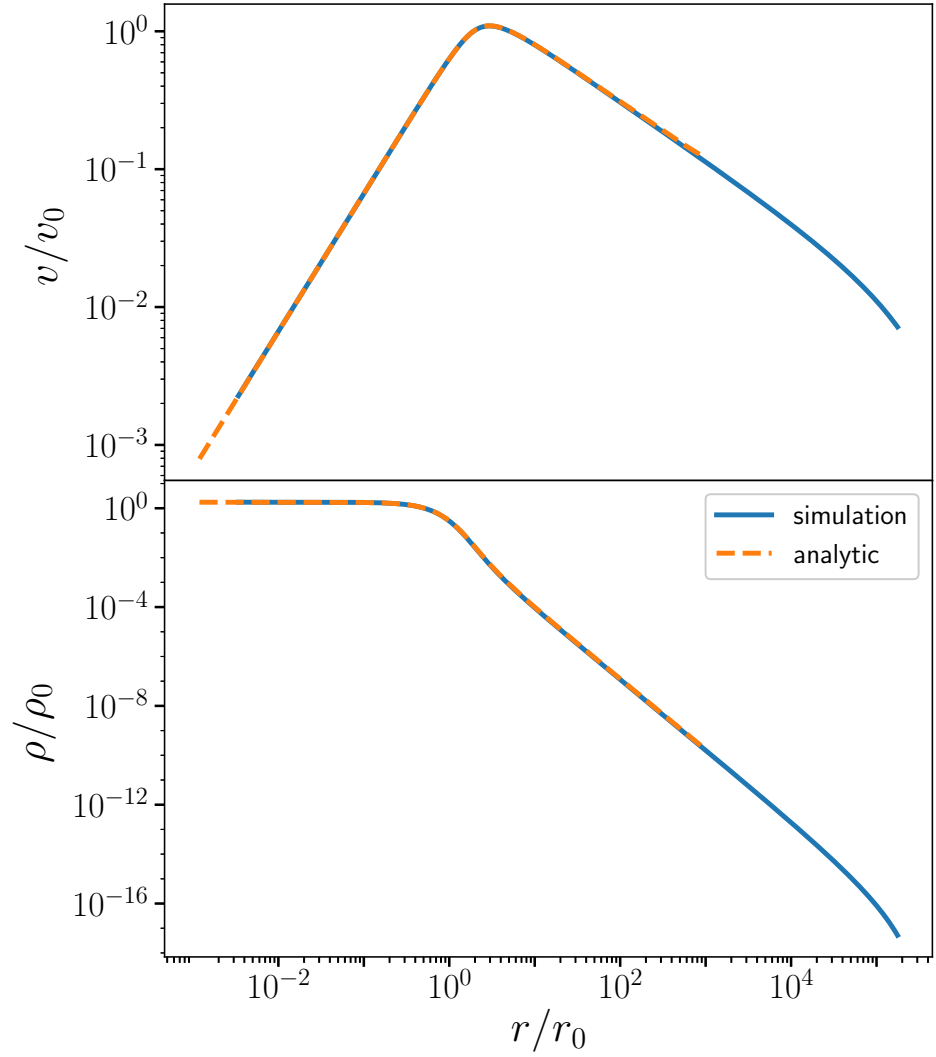}
    \caption{Simulated (blue) infall velocity (top panel) and density (bottom panel) during the simulation of the collapse of an ideal gas polytrope with an adiabatic constant $\gamma=1.3$ as a function of distance from the center, compared to the analytic self-similar solution (red). All quantities are dimensionless, where the values given by the simulation are normalized by the self-similar scaling.}
    \label{fig:analitic}
\end{figure}
%%%%%%%%%%%%%%%%%%%%%%%%%%%%%%%%%%%%%%%%%%%%%%%%%%

% Don't change these lines
%\bsp	% typesetting comment
\label{lastpage}
\end{document}